\documentclass[11pt]{article}
\usepackage[a4paper,margin=23mm,includefoot]{geometry}
\usepackage[T1]{fontenc}
\usepackage[utf8]{inputenc}
\usepackage{lmodern,microtype}
\usepackage{amsmath,amssymb,booktabs,array,multirow,graphicx}
\usepackage[numbers,sort]{natbib}
\usepackage[colorlinks=true,linkcolor=black,citecolor=black,urlcolor=blue]{hyperref}
\usepackage{caption,titlesec,needspace}
\titlespacing*{\section}{0pt}{2.2ex plus .5ex minus .2ex}{.85ex}
\titlespacing*{\subsection}{0pt}{1.65ex plus .4ex minus .2ex}{.55ex}

\providecommand{\doi}[1]{\href{https://doi.org/#1}{doi:\nolinkurl{#1}}}
\hypersetup{pdfauthor={Azzeddine Ihsine and Sara Ihsine},pdftitle={When Does Combining Software-Effort Estimates Help? Evidence from Four Public Project Datasets},pdfsubject={Empirical software effort estimation and forecast combination}}
\title{\textbf{When Does Combining Software-Effort Estimates Help?}\\Evidence from Four Public Project Datasets}
\author{Azzeddine Ihsine \quad Sara Ihsine\\[2pt]\small Inovionix\\[2pt]\small \texttt{azzeddine.ihsine@inovionix.com, sara.ihsine@inovionix.com}}
\date{}
\begin{document}
\maketitle
\begin{abstract}
Combining estimates can reduce individual prediction errors, but an ensemble may also dilute an unusually accurate estimator or repeat errors shared by its members. We study these competing effects on four archival software-effort datasets (Kitchenham: 145 projects, Desharnais: 77, China: 499, Maxwell: 63) with 100 repeated 70/30 training and test partitions per dataset. Three conventional models (parametric regression, analogy, and random forest) are combined with each other and, in Kitchenham, with a recorded human estimate. Out-of-fold calibration errors drive five combination rules: equal weighting, inverse-MSE weighting, a nonnegative simplex combination, signed Bates-Granger weighting, and Bates-Granger with covariance shrinkage. All preprocessing is learned within training folds. Model-error correlations range from 0.54 to 0.97. In Kitchenham, the human estimate dominates: equal weighting raises mean test log-MAE from 0.239 to 0.455, a loss that remains significant under a corrected resampled $t$-test with Holm adjustment, whereas the simplex rule limits the increase to 0.250. In Desharnais, China, and Maxwell, every rule lowers mean log-MAE (by 1.3 to 8.6\%) and BG with shrinkage beats the calibration-selected individual on 75, 83, and 83 of 100 splits, but none of these gains is statistically distinguishable from zero once the overlap between repeated splits is accounted for. On a completion-year split of Desharnais, equal weighting becomes slightly worse and inverse-MSE is near the individual baseline, while simplex and shrinkage retain mean gains of about 5\%. Combining correlated effort estimates is therefore a conditional decision: the harm of naive pooling next to a dominant source is clear, while the benefit elsewhere is small and uncertain.
\end{abstract}

\section{Introduction}
Decisions about software effort are routinely based on multiple imperfect signals: project history, a human judgment, analogy with similar projects, a simple parametric relation, or a learned model. Combining those signals is attractive because independently imperfect estimates may offset one another. But independence is a substantive assumption. Models trained on overlapping records and features may make similar mistakes, and two estimators with different labels are not necessarily independent sources of information. If one estimate is demonstrably more accurate than the others, averaging it with weaker alternatives can make the final estimate worse.

Research on software estimation emphasizes both the context dependence of predictive techniques and the difficulty of evaluating their performance fairly~\cite{jorgensen2004expert,jorgensen2007review,shepperd2012evaluating}. An updated review of ensemble effort estimation documents diverse combinations of learners and evaluation choices, without establishing a universally preferred rule~\cite{cabral2023ensemble}. Kocaguneli, Menzies and Keung report broad gains from ensembles over single learners across multiple datasets and error measures~\cite{kocaguneli2012ensemble}, whereas Minku and Yao examine the role of ensemble diversity, locality and multi-objective learning~\cite{minku2013locality,minku2013multiobjective}. These are central empirical precedents. Our contribution focuses more narrowly on numerical forecast-combination policies fitted from out-of-fold error second moments, including a setting where an independently recorded human estimate dominates. More recent work has pursued heterogeneous ensembles that jointly address missing data, feature selection, and adaptive combination~\cite{ali2024novel}, and multi-step dynamic ensemble selection on discretized effort targets~\cite{jadhav2024multistep}. Reviews of neural approaches and AI-based estimation continue to identify heterogeneous evidence, data quality, and limited industrial validation as central difficulties~\cite{boujida2025neural,santos2026ai}. These results motivate the use of ensembles, but do not answer whether combining \emph{already calibrated numerical estimates} is beneficial when their errors are correlated or one source is much more accurate than the others.

The forecast-combination literature explains this distinction mathematically. Individual error magnitudes are insufficient to determine the best weights because cross-estimator second moments, calibration-sample size, and regularization also affect loss~\cite{bates1969combination,clemen1989combining,timmermann2006forecast,smith2009puzzle}. We therefore compare forecast-combination \emph{policies}, measured on paired held-out project partitions, under differing marginal accuracy and error dependence, rather than develop a new learning architecture.

Our earlier D-POAF article studied uncertainty, score sensitivity, estimate aggregation, and downstream release decisions under synthetic experimental assumptions and noted that real project histories were still required~\cite{prior2026}. The present study addresses a narrower question on observed project records: \emph{when does combining effort estimates improve out-of-sample predictive accuracy, and when does it harm it?} It does not evaluate D-POAF's BVS, ERS, PVS, portfolio selection or Optimal Development Path (ODP) rules.

The contribution is empirical and comparative. We (i) measure dependence among out-of-fold error signals, (ii) compare five deployable combination policies with a calibration-selected individual on identical training/test splits, with a resampled test that accounts for overlapping training sets, (iii) report how often a rule helps rather than means alone, together with a time-ordered sensitivity check, and (iv) identify a dominant-estimator failure mode that makes the choice of combination policy consequential.

\section{Related Work and Research Positioning}
\subsection{Estimating software-project effort}
Expert judgment and statistical estimation differ in their information bases and failure modes. J\o rgensen's review and evidence-based uncertainty guidelines stress learning from recorded project performance without assuming that predictive evidence transfers automatically between organizations~\cite{jorgensen2004expert,jorgensen2005uncertainty}. Dataset and target construction affect reported accuracy, and Shepperd and MacDonell advocate evaluations aligned with the prediction task~\cite{shepperd2012evaluating}. A review of ensemble effort estimation documents many proposed combinations while also exposing methodological and contextual heterogeneity~\cite{cabral2023ensemble}.

Recent studies approach the problem from different directions. Ali, Ren, and Wu propose a composite framework incorporating missing-value imputation, feature selection, and a heterogeneous ensemble with adaptive combination~\cite{ali2024novel}. Jadhav and colleagues use multi-step dynamic ensemble selection on the \emph{same four named effort datasets} examined here, but discretize numeric effort into ranges and assess classifiers~\cite{jadhav2024multistep}. Their classification accuracy is not commensurable with our continuous held-out log-MAE or log-RMSE, and no head-to-head comparison is claimed. Boujida, Amazal, and Idri review neural-network estimation studies and point to the limited industrial uptake of these models~\cite{boujida2025neural}. Santos and Martins synthesize AI-based cost, effort, and schedule estimation research, highlighting data quality, reproducibility, and practical deployment challenges~\cite{santos2026ai}. These publications strengthen the case for context-specific testing. Our study isolates \emph{combination policy} rather than proposing a new imputation, feature-selection, classification, or neural architecture.

\subsection{Combining correlated predictions}
Consider a vector of forecast errors $\mathbf e=(e_1,\ldots,e_K)^\top$ and a weight vector $\mathbf w$ summing to one. Squared error of their weighted combination has expectation
\begin{equation}
\mathbb{E}[(\mathbf w^\top\mathbf e)^2]=\mathbf w^\top M\mathbf w,\quad M=\mathbb{E}[\mathbf e\mathbf e^\top].
\label{eq:second-moment}
\end{equation}
The diagonal of $M$ describes each estimator's mean-squared error and the off-diagonal elements describe shared error. The uncentred moment incorporates both stochastic covariance and bias interactions. Inverse-MSE weighting uses only the diagonal. A covariance-aware combination uses all of $M$, subject to the difficulty of estimating it from a finite calibration sample~\cite{bates1969combination,clemen1989combining,timmermann2006forecast}. Constrained nonnegative weights prevent extrapolative subtraction, whereas signed weights may cancel common errors but can be unstable. Shrinkage can improve conditioning while introducing its own bias toward more balanced weights~\cite{ledoit2004shrinkage,smith2009puzzle}.

\subsection{Relation to the earlier D-POAF study}
The earlier article~\cite{prior2026} used synthetic experiments to characterize prioritization and estimation under controlled error processes. Its results cannot be read as a test of the four-dataset combination results below, and none of them is reused here. We cite it for motivation. The present study shares authors and part of its methodological ancestry with that work, so it is an extension of the same research programme rather than an independent replication.

\section{Research Questions}
\begin{description}
\item[RQ1 (error dependence):] How strongly are out-of-fold errors correlated among the models, and how does that compare with an independently recorded human estimate?
\item[RQ2 (conditional accuracy):] Which combination rules outperform a calibration-selected individual under held-out log-MAE and log-RMSE, and are the differences distinguishable from resampling noise?
\item[RQ3 (stability of benefit):] On how many of the 100 repeated partitions does each combination help or harm, and how do signed weights and bias correction change the picture?
\item[RQ4 (practical boundary):] Does the pattern hold when later projects are predicted from earlier ones, and what would an organization need to test before relying on a combined estimate?
\end{description}
The split-level win counts of RQ3 are descriptive, whereas the inferential statement of RQ2 rests on the corrected test of Section~\ref{sec:estimands}.

\section{Data and Experimental Design}
\subsection{Four archival datasets and scope of prediction}
The study uses public files distributed through the DASE repository from the PROMISE tradition~\cite{promise2005,dase}. After removing incomplete records, the data contain 145 Kitchenham projects, 77 Desharnais projects, 499 China projects, and 63 Maxwell projects. The Kitchenham dataset includes a historical first estimate, making it the only one of the four with a separately recorded human forecast~\cite{kitchenham2002empirical}. Desharnais and Maxwell have distinct historical origins~\cite{desharnais1989,maxwell2002applied}. Each record is a completed project. No sprint-level data or D-POAF Waves~\cite{dpoafcanonical2026} are used.

The common target is the natural logarithm of actual effort. For a strictly positive estimate $\widehat y$ and actual $y$, the prediction error is $e=\log\widehat y-\log y$, which treats equal multiplicative over- and under-estimation symmetrically. For example, $|e|=0.1$ corresponds to a multiplicative deviation $e^{0.1}\approx1.105$. We report test log-MAE and log-RMSE, the latter aligned with methods fitted by squared-error minimization. We do not rely on MMRE, whose behavior is known to be problematic~\cite{foss2003mmre}.

\begin{table}[t]
\centering\small
\caption{Datasets and prediction sources, after removal of incomplete records.}
\label{tab:datasets}
\begin{tabular}{lrrp{6.65cm}}
\toprule
Dataset & Projects & Sources & Predictors \\
\midrule
Kitchenham & 145 & 4 & Recorded first estimate (human) plus three models on adjusted function points.\\
Desharnais & 77 & 3 & Team and manager experience, transactions, entities, adjusted function points, adjustment factor, language (nominal), without completion year.\\
China & 499 & 3 & Function-point components, resource level, development type (nominal), adjustment factor, without recorded duration.\\
Maxwell & 63 & 3 & Size, application and environment attributes, technical factors t01 to t15, without recorded duration.\\
\bottomrule
\end{tabular}
\end{table}

\subsection{Estimator inputs and preprocessing}
We evaluate three model families on each dataset. The first is an ordinary-least-squares regression of log effort on log size. The second is an analogy estimator using three nearest neighbours on standardized features. The third is a random forest with 150 trees and minimum leaf size two~\cite{breiman2001rf,pedregosa2011sklearn}. Kitchenham additionally contributes the recorded human first estimate, which is used as recorded and never fitted.

Predictors are restricted to attributes plausibly known when a project is estimated. Recorded actual duration is excluded in China and Maxwell, and the Desharnais completion year (\texttt{YearEnd}) is excluded from all predictors and used only to form the time-ordered split of Section~\ref{sec:design}. Every preprocessing decision is learned from the training rows of the fit at hand, in the outer 70/30 split and in each of the five calibration folds: nominal attributes (Desharnais language, China development type) are one-hot encoded with the categories observed in training, unseen test categories map to all-zero indicators, and a nonnegative attribute is log-transformed with $\log(1+x)$ when its training maximum exceeds 20. Standardization for the analogy estimator is fitted on training rows only. The target is never used in preprocessing.

\subsection{Partitioning, calibration, and held-out evaluation}\label{sec:design}
For each dataset, 100 seeded random permutations reserve 30\% of projects as a test set and 70\% as a training/calibration set. Within the calibration set, each model's error is computed through five-fold out-of-fold predictions. Models fitted on the whole 70\% then predict the held-out 30\%. Every combination policy receives the same calibration errors and is evaluated on the same test projects within a repetition. The 100 partitions reuse the same projects, so they are not 100 independent replications, and Section~\ref{sec:estimands} accounts for this.

Let $E_c\in\mathbb{R}^{m\times K}$ be the matrix of out-of-fold calibration errors, with rows representing projects and columns estimators. The fitted uncentred second-moment matrix is $\widehat M=E_c^\top E_c/m$. We evaluate the following established forecast-combination policies:
\begin{align}
\text{Equal:}\quad &w_i=1/K,\nonumber\\
\text{Inverse-MSE:}\quad &w_i\propto1/\widehat M_{ii},\nonumber\\
\text{Simplex:}\quad &\mathbf w_{\mathrm{simplex}}=\operatorname*{arg\,min}_{\mathbf w\geq0,\,\mathbf1^\top\mathbf w=1}\mathbf w^\top\widehat M\mathbf w,\label{eq:weights}\\
\text{BG-style signed:}\quad &\mathbf w=\frac{\widehat M^{-1}\mathbf1}{\mathbf1^\top\widehat M^{-1}\mathbf1},\nonumber
\end{align}
with a fifth rule replacing $\widehat M$ by a Ledoit-Wolf shrinkage estimate toward a scaled identity, fitted with the errors treated as uncentred, before applying the same signed-weight formula~\cite{ledoit2004shrinkage}. These covariance-aware signed weights are a Bates-Granger-style extension to $K$ possibly biased forecasts, not a new method~\cite{bates1969combination,timmermann2006forecast}. They solve the sum-to-one constrained minimum-MSE problem when $\widehat M$ is positive definite (Lagrange condition $2\widehat M\mathbf w=\lambda\mathbf 1$). A diagonal ridge of $10^{-12}I$ stabilizes the solve when the matrix is nearly singular. The baseline is the \emph{best individual selected on calibration MSE}, not the estimator that turns out best on the test set, which would be an unavailable oracle.

A bias-corrected sensitivity variant subtracts each estimator's mean calibration error from calibration and test errors before fitting and scoring. We report it as a robustness analysis, not as an assumption that a future estimate's bias is stationary. No parameter was tuned on held-out test outcomes.

\paragraph{Time-ordered sensitivity check.} Random partitions let earlier projects be predicted from later ones. Desharnais is the only dataset with a time attribute: completion years 1983 (3 projects), 1984 (5), 1985 (22), 1986 (28), 1987 (14) and 1988 (5). We train and calibrate on the 58 projects completed in 1983 to 1986 and test on the 19 completed in 1987 to 1988, with the same pipeline. Because this partition is fixed, we repeat it over 50 seeds of the calibration folds and of the random forest, which vary the out-of-fold errors, the selected individual and the fitted weights. Completion year is a proxy for chronology: start dates are not recorded, so a training project may have overlapped in time with a test project. Maxwell contains no documented date or delivery order, and no time-ordered evaluation is attempted for it, for China, or for Kitchenham.

\subsection{Estimands and statistical analysis}\label{sec:estimands}
For method $m$ on split $s$ of dataset $d$, define held-out loss $L_{dsm}$ (log-MAE or log-RMSE). We report the mean $\overline L_{dm}$ across 100 splits and the descriptive ratio-of-means reduction
\begin{equation}
G_{dm}=100\left(1-\frac{\overline L_{dm}}{\overline L_{d,\mathrm{best}}}\right),
\label{eq:gain}
\end{equation}
where positive values favor combination. We also count $W_{dm}=\sum_s\mathbf1\{L_{dsm}<L_{ds,\mathrm{best}}\}$. Exact ties (4 of 2{,}000 comparisons) are not counted as wins.

Because training sets overlap across repetitions, the naive paired $t$-test on the 100 split-level differences understates variance. We use the corrected resampled $t$-test of Nadeau and Bengio~\cite{nadeau2003inference}: with $\delta_s=L_{ds,\mathrm{best}}-L_{dsm}$, $J=100$ and test/training ratio $n_2/n_1$,
\begin{equation}
t=\frac{\bar\delta}{\sqrt{\left(\frac1J+\frac{n_2}{n_1}\right)\hat\sigma^2_\delta}},
\label{eq:nb}
\end{equation}
compared with a $t$ distribution with $J-1$ degrees of freedom, and the corresponding 95\% interval for $\bar\delta$. The 20 tests (five rules, four datasets) on log-MAE are adjusted with Holm's procedure~\cite{holm1979simple}. The correction is known to be conservative. It addresses the dependence between repetitions on the same projects, but it does not turn a fixed dataset into a sample of organizations, so no pooled meta-estimate across the four datasets is computed.

\section{Results}
\subsection{RQ1: model errors are strongly correlated}
\begin{figure}[t]
\centering\includegraphics[width=.99\linewidth]{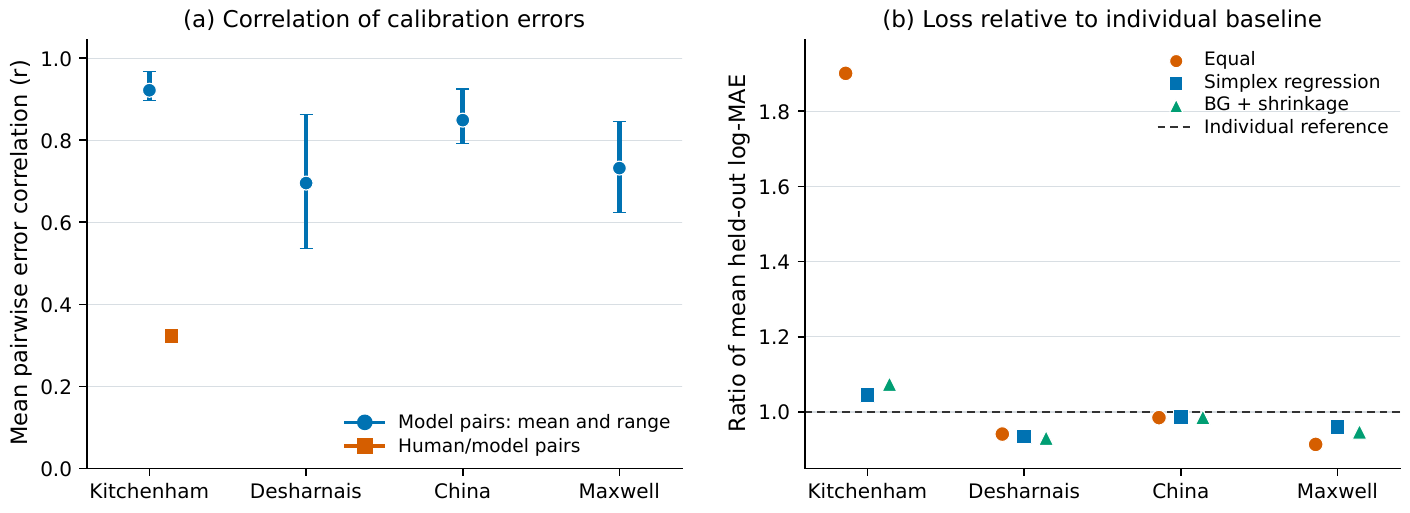}
\caption{(a) Mean of the three pair-specific calibration-error correlations per dataset (circles) with their minimum-to-maximum range (bars, not a confidence interval). Kitchenham also has three human-to-model pairs (squares). (b) Ratio of the mean held-out log-MAE of a combination rule to that of the calibration-selected best individual over 100 splits. Values below one favor combination.}
\label{fig:profiles}
\end{figure}

Table~\ref{tab:correlations} and Figure~\ref{fig:profiles}(a) report the mean within-split Pearson correlations of calibration errors. Among model-based estimators, the correlations range from 0.54 to 0.97, consistent with overlapping errors associated with shared predictors and training populations. Kitchenham makes the contrast visible. Its model-to-model correlation averages 0.92, whereas correlations of the recorded human first estimate with model errors are only 0.31 to 0.34. Low error correlation does not prove informational independence, but it shows that the recorded human estimate has a less-correlated error signal in this dataset.

\begin{table}[t]
\centering\small
\caption{Calibration-error Pearson correlations, averaged over 100 repeated splits. The range is over the three model pairs in each dataset. Human pairs exist only for Kitchenham.}
\label{tab:correlations}
\begin{tabular}{lccc}\toprule
Dataset & Model-pair range & Model-pair mean & Human-to-model range \\\midrule
Kitchenham & 0.90 to 0.97 & 0.92 & 0.31 to 0.34 \\
Desharnais & 0.54 to 0.86 & 0.70 & n/a \\
China & 0.79 to 0.92 & 0.85 & n/a \\
Maxwell & 0.62 to 0.85 & 0.73 & n/a \\
 
\bottomrule\end{tabular}
\end{table}

\subsection{RQ2: combining helps on average in three datasets and harms in one}
Table~\ref{tab:accuracy} gives the held-out mean losses and Table~\ref{tab:improvement} the relative changes. Under log-MAE, equal weighting scores 0.403, 0.732, and 0.504 on Desharnais, China, and Maxwell, against 0.428, 0.743, and 0.552 for the calibration-selected individual. Every combination rule has a lower mean loss than the individual in these three datasets, by 5.8 to 6.9\% in Desharnais, 1.3 to 1.6\% in China and 2.8 to 8.6\% in Maxwell, and the same holds for log-RMSE. Differences among combination rules are small relative to these gains.

\begin{table}[t]
\centering\scriptsize
\caption{Held-out mean log-MAE and log-RMSE over 100 repeated 70/30 splits (identical splits for all methods, smaller is better). ``Best individual'' is selected on calibration errors.}
\label{tab:accuracy}
\setlength{\tabcolsep}{4pt}
\begin{tabular}{l*{8}{r}}
\toprule
&\multicolumn{4}{c}{log-MAE}&\multicolumn{4}{c}{log-RMSE}\\
\cmidrule(lr){2-5}\cmidrule(lr){6-9}
Method & Kitch. & Desh. & China & Maxw. & Kitch. & Desh. & China & Maxw.\\
\midrule
Best individual & 0.239 & 0.428 & 0.743 & 0.552 & 0.358 & 0.537 & 0.922 & 0.652 \\
Equal & 0.455 & 0.403 & 0.732 & 0.504 & 0.597 & 0.505 & 0.919 & 0.620 \\
Inverse-MSE & 0.318 & 0.402 & 0.731 & 0.508 & 0.417 & 0.501 & 0.916 & 0.617 \\
Simplex regression & 0.250 & 0.401 & 0.733 & 0.530 & 0.356 & 0.499 & 0.913 & 0.634 \\
Bates-Granger & 0.256 & 0.403 & 0.734 & 0.536 & 0.359 & 0.501 & 0.914 & 0.643 \\
BG + shrinkage & 0.257 & 0.399 & 0.733 & 0.523 & 0.356 & 0.496 & 0.913 & 0.628 \\
 
\bottomrule\end{tabular}
\end{table}

\begin{table}[t]
\centering\small
\caption{Relative change in mean test log-MAE versus the calibration-selected best individual (\%, Eq.~\ref{eq:gain}). Positive values mean lower error for the combination.}
\label{tab:improvement}
\begin{tabular}{lrrrr}\toprule
Method & Kitchenham & Desharnais & China & Maxwell\\\midrule
Equal & -90.0\% & +5.8\% & +1.5\% & +8.6\% \\
Inverse-MSE & -32.9\% & +6.2\% & +1.6\% & +7.9\% \\
Simplex regression & -4.5\% & +6.4\% & +1.4\% & +4.0\% \\
Bates-Granger & -7.0\% & +5.8\% & +1.3\% & +2.8\% \\
BG + shrinkage & -7.5\% & +6.9\% & +1.4\% & +5.2\% \\
 
\bottomrule\end{tabular}
\end{table}

Kitchenham is the counterexample. The recorded human first estimate has mean log-MAE 0.239 and log-RMSE 0.358, and is selected as the best individual in every split, while the three models are much less accurate. Pooling it with them is harmful. Equal weighting raises mean log-MAE to 0.455 (90\% above the baseline) and inverse-MSE weighting to 0.318 (33\% above). Simplex regression limits the penalty to 0.250, and its mean log-RMSE of 0.356 is level with the human estimate's 0.358.

Table~\ref{tab:stats} tests these differences with the corrected resampled $t$-test. One effect survives Holm adjustment: the loss of equal weighting in Kitchenham, $-0.215$ log-MAE (95\% interval $-0.296$ to $-0.134$). The loss of inverse-MSE weighting in Kitchenham has an interval that excludes zero ($p=0.009$) but does not survive the adjustment. None of the gains in Desharnais, China and Maxwell is distinguishable from zero: all intervals contain it, with unadjusted $p$ between 0.17 and 0.54. Once the overlap between repeated training sets is taken into account, the variability of the paired differences is large relative to the average gain of combining. The evidence is therefore asymmetric: naive pooling next to a dominant source is clearly harmful, while the benefit of combining comparable models is consistent in sign but small and statistically uncertain on datasets of this size.

\begin{table}[t]
\centering\scriptsize
\caption{Mean reduction of test log-MAE relative to the calibration-selected individual over 100 splits, with 95\% intervals from the corrected resampled $t$-test (Eq.~\ref{eq:nb}). Positive values favor combination. $^{*}$: significant at 5\% after Holm adjustment over the 20 comparisons.}
\label{tab:stats}
\setlength{\tabcolsep}{3pt}
\begin{tabular}{lcccc}\toprule
Method & Kitchenham & Desharnais & China & Maxwell\\\midrule
Equal & -0.215 [-0.296, -0.134]$^{*}$ & +0.025 [-0.056, +0.106] & +0.011 [-0.015, +0.037] & +0.047 [-0.033, +0.127] \\
Inverse-MSE & -0.079 [-0.138, -0.020] & +0.027 [-0.047, +0.101] & +0.012 [-0.012, +0.036] & +0.043 [-0.019, +0.106] \\
Simplex regression & -0.011 [-0.030, +0.009] & +0.028 [-0.027, +0.082] & +0.010 [-0.006, +0.026] & +0.022 [-0.019, +0.063] \\
Bates-Granger & -0.017 [-0.038, +0.004] & +0.025 [-0.031, +0.081] & +0.009 [-0.007, +0.026] & +0.015 [-0.033, +0.064] \\
BG + shrinkage & -0.018 [-0.041, +0.006] & +0.030 [-0.030, +0.090] & +0.010 [-0.006, +0.027] & +0.029 [-0.018, +0.076] \\
 
\bottomrule\end{tabular}
\end{table}

\subsection{RQ3: average gains hide split-to-split failures}
\begin{figure}[t]
\centering\includegraphics[width=.88\linewidth]{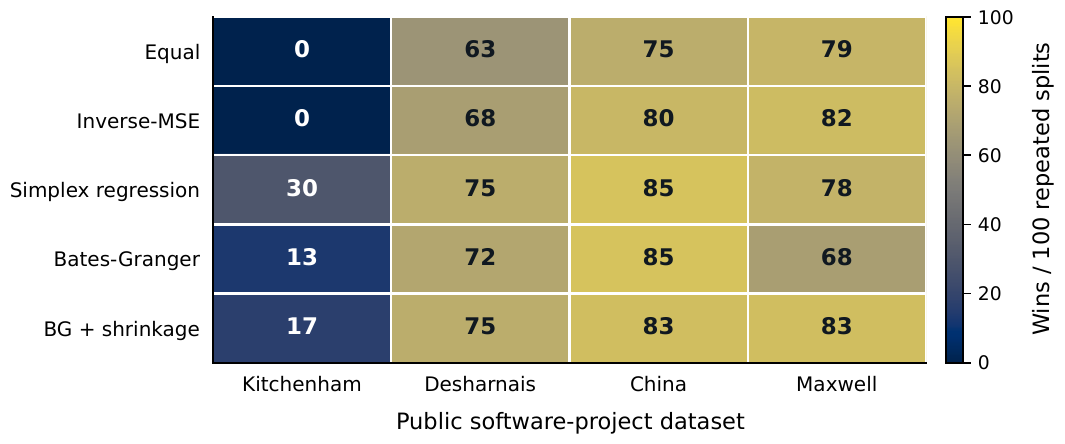}
\caption{Favorable held-out splits out of 100 per rule and dataset. Repetitions reuse projects, so these are not independent success trials.}
\label{fig:wins}
\end{figure}

Table~\ref{tab:wins} and Figure~\ref{fig:wins} count, out of 100 repetitions, the splits on which a rule has strictly lower test log-MAE than the calibration-selected individual. BG with shrinkage wins on 75 Desharnais, 83 China and 83 Maxwell splits, but on only 17 Kitchenham splits, and equal weighting never wins in Kitchenham. In the three datasets where combining helps on average, every rule still loses on 15 to 37 splits. These counts describe repeated partitions of the same projects and are not probabilities of success in a future deployment, but they show that a favorable mean coexists with a substantial share of adverse cases.

\begin{table}[t]
\centering\small
\caption{Wins out of 100 matched random partitions against the calibration-selected best individual on test log-MAE.}
\label{tab:wins}
\begin{tabular}{lrrrr}\toprule
Method & Kitchenham & Desharnais & China & Maxwell\\\midrule
Equal & 0 & 63 & 75 & 79 \\
Inverse-MSE & 0 & 68 & 80 & 82 \\
Simplex regression & 30 & 75 & 85 & 78 \\
Bates-Granger & 13 & 72 & 85 & 68 \\
BG + shrinkage & 17 & 75 & 83 & 83 \\
 
\bottomrule\end{tabular}
\end{table}

The signed covariance-aware policies sometimes assign negative weights, which is algebraically legitimate but a governance concern when users expect a convex average. At least one negative weight occurs in 99 Kitchenham, 44 Desharnais, 12 China, and 51 Maxwell Bates-Granger fits, and shrinkage reduces these counts to 94, 4, 7, and 16 (Table~\ref{tab:neg}). Shrinkage does not guarantee nonnegative weights, whereas the simplex constraint does. Mean weight on the human source in Kitchenham is 0.90 for simplex, 0.89 for Bates-Granger, 0.85 for shrinkage and 0.61 for inverse-MSE, against 0.25 under equal weighting, which explains the ordering of their losses in the dominant-estimator setting.

\begin{table}[t]
\centering\small
\caption{Splits (of 100) with at least one negative fitted weight.}
\label{tab:neg}
\begin{tabular}{lrrrr}\toprule
Method & Kitchenham & Desharnais & China & Maxwell\\\midrule
Bates-Granger & 99 & 44 & 12 & 51 \\
BG + shrinkage & 94 & 4 & 7 & 16 \\
 
\bottomrule\end{tabular}
\end{table}

With mean-bias correction the ordering does not change (Appendix, Table~\ref{tab:bias}): shrinkage yields mean test log-MAE 0.255 in Kitchenham versus 0.241 for the corrected best individual, while Desharnais has 0.399 versus 0.427 and China 0.732 versus 0.744.

\subsection{RQ4: a time-ordered check and the practical boundary}
Table~\ref{tab:chrono} reports the Desharnais completion-year split. Analogy is the most accurate single model on the 1987 to 1988 projects (log-MAE 0.343) and is selected on calibration error in 46 of 50 seeds. In the other four the selection picks the parametric or random-forest model, which raises the mean loss of the selected individual to 0.355. Against it, equal weighting is slightly worse ($-1.6$\%, 4 wins of 50) and inverse-MSE weighting about level (7 wins), whereas the simplex rule (0.334, $+5.7$\%, 41 wins) and shrinkage (0.338, $+4.8$\%, 37 wins) keep gains similar to those of the random partitions. The test set holds 19 projects, so the comparison is indicative only. It agrees with the random-split results in one respect, that rules which can down-weight a weak member are safer than rules which cannot, and it shows that the advantage of naive pooling observed on random splits need not carry over to later projects.

\begin{table}[t]
\centering\small
\caption{Desharnais, trained and calibrated on projects completed in 1983 to 1986 (58), tested on 1987 to 1988 (19). Mean test log-MAE over 50 seeds of calibration folds and random forest, its range, relative change against the calibration-selected individual, and seeds (of 50) on which the rule beats it. Lower block: the three models alone.}
\label{tab:chrono}
\begin{tabular}{lcccc}\toprule
Method & Mean log-MAE & Range & Change & Wins / 50\\\midrule
Best individual & 0.355 & 0.343 to 0.562 & +0.0\% & n/a \\
Equal & 0.360 & 0.353 to 0.367 & -1.6\% & 4 \\
Inverse-MSE & 0.352 & 0.337 to 0.367 & +0.8\% & 7 \\
Simplex regression & 0.334 & 0.305 to 0.397 & +5.7\% & 41 \\
Bates-Granger & 0.355 & 0.306 to 0.422 & -0.2\% & 25 \\
BG + shrinkage & 0.338 & 0.309 to 0.371 & +4.8\% & 37 \\
\midrule
Parametric alone & 0.562 & n/a & & \\
Analogy alone & 0.343 & n/a & & \\
ML alone & 0.401 & 0.373 to 0.426 & & \\
 
\bottomrule\end{tabular}
\end{table}

Two mechanisms explain the overall pattern. \emph{Error diversification} is possible when sources have comparable marginal accuracy and do not reproduce all of one another's failures. \emph{Dilution of a dominant source} occurs when averaging adds less accurate estimates, even when those are correlated with one another rather than with the best source. Covariance-aware and constrained weights respond to both mechanisms, but estimation noise makes fitted weights unstable~\cite{smith2009puzzle,ledoit2004shrinkage}. Before relying on a combined estimate, an organization would need to measure held-out performance on chronologically later projects of its own, compare against retaining its strongest individual source, document input availability, and monitor deterioration after adoption.

\section{Discussion}
\subsection{Where the contribution lies}
The principal contribution is the conjunction of real residual-correlation evidence, a calibrated set of classical combination rules, their contrasting out-of-sample behavior, and an inference that accounts for the dependence between repeated splits. In contrast with work emphasizing heterogeneous model construction or dynamic classification~\cite{ali2024novel,jadhav2024multistep}, we hold the prediction channels fixed within each split and ask when combining them helps. Because training protocols, targets, and metrics differ, we do not rank our methods against results reported in those studies.

Correlation alone is not a success criterion. Kitchenham has low human-to-model error correlation, yet the human estimate is so strong that adding weaker predictions hurts. A rule's value depends on both \emph{diversity} and \emph{relative marginal accuracy}. The corrected tests add a second qualification to the common recommendation to use ensembles~\cite{kocaguneli2012ensemble}: on datasets of 63 to 499 projects, average gains of a few percent are within resampling noise, and many reported ensemble advantages evaluated with naive repeated-split statistics may be similarly fragile. Only the harm of naive pooling next to a dominant source is established with confidence here.

\subsection{Implications for governed decision support}
A decision framework such as D-POAF can record not only an aggregated effort estimate but also which sources contributed, how their errors were calibrated, which policy generated the weights, the calibration-sample size, and whether signed weights were allowed. The results motivate a conservative decision procedure: use the strongest calibration-selected individual as a benchmark, examine constrained (simplex) and shrunk alternatives, and adopt a combination only if it passes a held-out test on later projects. No rule is uniformly preferable across all settings. These are audit requirements suggested by the evidence, not claims that any platform has reduced engineering effort. The results do not measure project selection, sequencing, post-release value, or delivery time, and random-forest estimates are not evidence about generative-AI forecasting.

\subsection{Replication}
An exact rerun requires the original dataset files, the preprocessing and combination rules, and the random-number-generation scheme used to reconstruct the split and fold assignments and model seeds. The restricted verification package retains the research scripts, dataset snapshots and split-level loss summaries, and can be used to regenerate predictions under a compatible environment. It does not contain precomputed project-level prediction vectors or enumerated split indices. The public package contains aggregate tables, published test statistics, win counts and figure data. It does not contain the paired split-level losses needed to recompute the corrected tests independently.

\section{Threats to Validity}
\textbf{Construct validity.} The target is historical project-level effort recorded in archival data. Log-scale error captures multiplicative inaccuracy but is not equivalent to monetary cost or delivered value. The four datasets use different project definitions, effort units and attribute sets. Kitchenham's first estimate may have been produced under organizational conditions not comparable to the three model predictions.

\textbf{Internal validity.} The same projects appear in many of the 100 repeated splits. The corrected resampled test addresses this dependence but is approximate and conservative. The baseline is selected on calibration MSE although log-MAE is the primary endpoint, which can disadvantage MAE-specific selections. All preprocessing is learned within training folds and post-hoc attributes (duration, completion year) are excluded, but the availability of each remaining predictor at the original estimation date cannot be verified from the archives. The Desharnais time-ordered check relies on completion year, not start date. Maxwell's technical attributes t01 to t15 are not timestamps, and no chronological evaluation is claimed for Maxwell, China or Kitchenham. Duplicate projects and collection biases in the archives are not independently validated.

\textbf{External validity.} Four archival datasets do not represent present-day industrial projects in general. Random 70/30 splits test generalization within each historic dataset, not forecasting for later projects, other companies, or changing practices, and the single time-ordered check covers 19 test projects. Only one dataset records a human estimate and none records a generative-AI estimate, so conclusions about human-AI teams are out of scope.

\textbf{Analytical choices and multiplicity.} Five rules, two losses and four datasets are compared. Inference is restricted to log-MAE with Holm adjustment over 20 comparisons, while log-RMSE results and win counts are descriptive. The authors develop the broader D-POAF framework, so negative and null findings are reported in full to limit confirmation bias.

\section{Conclusion}
Across four archival software-project datasets, combining effort estimates is a conditional decision. The three model-based error streams are strongly correlated. In Kitchenham an independently recorded human estimate is markedly stronger than the models, and equal weighting nearly doubles its error, the one effect that survives a corrected, multiplicity-adjusted test. In Desharnais, China, and Maxwell every combination rule lowers mean log-MAE by 1 to 9\%, but these gains are within resampling noise, and each rule still loses on 15 to 37 of 100 splits. On the completion-year split of Desharnais, the largest favorable mean differences are observed for simplex and shrinkage, while inverse-MSE is near the individual reference and equal weighting is slightly worse. These findings favor benchmarking constrained and shrunk alternatives against a strong individual source rather than assuming that aggregation will help. The next decisive validation would use chronologically ordered projects from other organizations and, if human-AI claims are intended, actual AI forecasts and human decisions.

\section*{Availability and Reproducibility}
The four datasets are public and were retrieved from the DASE repository (commit 37945fb, CC0 1.0)~\cite{dase,promise2005}. The accompanying aggregate-results archive was published on Zenodo on 3 November 2026 (version 1.0, \href{https://doi.org/10.5281/zenodo.23913209}{doi:10.5281/zenodo.23913209})~\cite{effort_aggregates2026}. It contains method-level mean losses with their split-level standard deviations, calibration-error correlation aggregates, split-level win counts, negative-weight counts, the corrected test statistics, the time-ordered results, and the data behind each figure. The earlier Zenodo record~\cite{package2026} belongs to the simulation-based predecessor and does not contain this study. The restricted verification package contains research scripts, source dataset snapshots, split-level losses and a deterministic procedure for regenerating the splits and fold assignments. It does not include stored project-level prediction vectors. Access to the restricted package can be requested from the corresponding author under applicable data and rights conditions. Neither package contains D-POAF platform or production code. The analysis uses Python 3.13, NumPy 2.5.3, SciPy 1.18.1, pandas 3.0.5 and scikit-learn 1.9.1.

\section*{Competing Interests}
The authors contribute to D-POAF and are affiliated with Inovionix, which develops software implementing parts of that framework. All empirical comparisons use independent classical estimators, and negative outcomes for combination are reported. Independent replication is desirable. The authors declare no other competing interests.

\bibliographystyle{unsrtnat_hyphenated}
\bibliography{refs}
\clearpage
\appendix
\section{Additional Analytical Details}
\subsection{Losses and relative gains}
For each split, the absolute error is $|\widehat{\log y}-\log y|$ and the squared error is $(\widehat{\log y}-\log y)^2$. Log-MAE averages absolute errors across test projects, and log-RMSE is the square root of the mean squared error in that test set. The arithmetic mean of the 100 split-level values is reported, not a pooled error over all test projects. The descriptive standard deviation over repetitions does not identify between-organization uncertainty.

\subsection{Bias correction sensitivity}
The calibration-mean adjustment uses only training-side errors. Table~\ref{tab:bias} gives held-out log-MAE of the adjusted predictions. The main directional patterns are unchanged: Kitchenham retains its accurate human estimate, and all combinations remain better on average in the other three datasets. This descriptive check cannot account for future changes in estimator bias.
\begin{table}[h]
\centering\small
\caption{Sensitivity: mean log-MAE after subtracting each estimator's calibration mean error (100 repeated splits).}
\label{tab:bias}
\begin{tabular}{lrrrr}\toprule
Method & Kitchenham & Desharnais & China & Maxwell\\\midrule
Best individual & 0.241 & 0.427 & 0.744 & 0.555 \\
Equal & 0.455 & 0.405 & 0.731 & 0.507 \\
Inverse-MSE & 0.317 & 0.402 & 0.730 & 0.511 \\
Simplex regression & 0.250 & 0.400 & 0.733 & 0.531 \\
Bates-Granger & 0.254 & 0.403 & 0.733 & 0.539 \\
BG + shrinkage & 0.255 & 0.399 & 0.732 & 0.525 \\
 
\bottomrule\end{tabular}
\end{table}

\end{document}